\documentclass[twocolumn,superscriptaddress,showpacs,preprintnumbers,amsmath,amssymb,aps,prb]{revtex4}
\usepackage{graphicx}
\usepackage{dcolumn}
\usepackage{bm}
\begin{document}
\title{Natural Time Analysis of Seismicity in California: The epicenter of an impending mainshock}
\author{P. A. Varotsos}
\thanks{{\bf Correspondence to:} P. Varotsos (pvaro@otenet.gr)}
\affiliation{Solid State Section and Solid Earth Physics
Institute, Physics Department, University of Athens,
Panepistimiopolis, Zografos 157 84, Athens, Greece}
\author{N. V. Sarlis}
\affiliation{Solid State Section and Solid Earth Physics
Institute, Physics Department, University of Athens,
Panepistimiopolis, Zografos 157 84, Athens, Greece}
\author{E. S. Skordas}
\affiliation{Solid State Section and Solid Earth Physics
Institute, Physics Department, University of Athens,
Panepistimiopolis, Zografos 157 84, Athens, Greece}


\begin{abstract}
Upon employing the analysis in a new time domain, termed natural
time, it has been recently demonstrated that a remarkable change
of seismicity emerges before major mainshocks in California. What
constitutes this change is that the fluctuations of the order
parameter of seismicity exhibit a clearly detectable minimum. This
is identified by using a natural time window sliding event by
event through the time series of the earthquakes in a wide area
and comprising a number of events that would occur on the average
within a few months or so. Here, we suggest a method to estimate
the epicentral area of an impending mainshock by an additional
study of this minimum using an area window sliding through the
wide area. We find that when this area window surrounds (or is
adjacent to) the future epicentral area, the minimum of the order
parameter fluctuations in this area appears at a date very close
to the one at which the minimum is observed in the wide area. The
method is applied here to major mainshocks that occurred in
California during the recent decades including the strongest one,
i.e., the 1992 Landers earthquake.
\end{abstract}

\maketitle

It may be considered (e.g. refs. \onlinecite{TUR97,HOL06}) that
earthquakes are non-equilibrium critical phenomena.  Their complex
correlations in time, space and magnitude $M$  have been
investigated by various procedures (e.g., see refs.
\cite{HUA08,TEL10B,LEN11,LIP12,TEN12}) including a very recent
development \cite{TEN12} of the complex networks approach. In this
study, we employ the analysis in a new time domain, termed natural
time\cite{NAT02,NAT02A,TAN04}, $\chi$, since it allows us to
identify\cite{PNAS} when a system approaches a critical point (for
a review see ref. \onlinecite{SPRINGER}).

Seismic Electric Signals (SES) are low frequency ($\leq 1$Hz)
electric signals \cite{VAR84A} preceding earthquakes. They are
emitted from the future focal region probably through the
following mechanism \cite{TECTO13}: When in the focal region the
stress reaches a {\em critical} value $\sigma_{cr}$, a cooperative
orientation of the electric dipoles (that have been formed due to defects\cite{VAR78,VAR80K133,VAR82,VAR82B,VAR08438}) occurs, which leads to the
emission of a transient electric signal that constitutes an SES.
Several such signals within a short time  are called SES activity
\cite{VAR96B,NAT03A,NAT03B}. By combining the SES physical
properties,  the epicenter and the magnitude  of the impending
mainshock  can be determined\cite{VAR96B}. Furthermore, by making
use of the small earthquakes subsequent to the initiation of an
SES activity and analyzing them in natural time (see below), the
occurrence time of an impending mainshock can be identified as
follows: We compute the value of the order parameter $\kappa_1$ of
seismicity (see below)  and then find\cite{NAT06B,SAR08} that a
mainshock occurs in a few days to one week after the $\kappa_1$
value  approaches 0.070. By applying this procedure, several
successful predictions have been issued \cite{SPRINGER} in Greece.

\begin{figure}[ht]
\begin{center}
\centerline{\includegraphics[width=.4\textwidth,angle=-90]{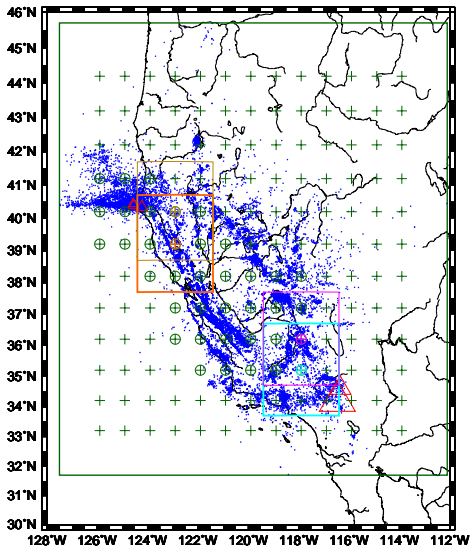}}
\caption{(color online) Map of the earthquake epicenters (blue
dots) in the wide area (green rectangle): $N_{31.7}^{45.7}
W_{127.5}^{112.1}$ ($M \geq 2.5$) during 1 January 1979 - 1
January 2004 as reported by NCEDC. The major mainshocks in
California discussed in this paper are shown with red triangles
and the neighboring rectangles depict their tentative epicentral
area estimated by means of the procedure described in the text. The
156 green plus symbols depict the locations of the area window of
size $3^o \times 3^o$ sliding through the wide area $14^o \times
15.4^o$ out of which 38 are encircled in order to show that they
contain at least 24 events occurring on average in three
months.}\label{fig1}
\end{center}
\end{figure}

 The challenging question raises, however,
what can we do when geoelectrical data are lacking which was the
case before the occurrence of the most strong earthquakes (EQs)
worldwide during the recent decades. It is the main objective of
this paper to attempt answering this question motivated by an
important fact which has  been recently reported \cite{TECTO13}: The
initiation of an SES activity is accompanied by a clearly
{detectable} change in seismicity which constitutes an independent
geophysical dataset of a different physical observable. In
particular, by analyzing in natural time the time series of
earthquakes, we found that the fluctuations of the order parameter
of seismicity  exhibited a clearly detectable minimum
approximately at the time of the initiation of the SES activity.
Furthermore, we showed \cite{TECTO13} that these two phenomena
(SES activity, minimum of the order parameter fluctuations of
seismicity) beyond their co-location in time they were also linked
closely in space. Here, we take advantage of these recent findings
 and show the following: Since the natural
time analysis of the seismic data alone can lead to the
identification of the minimum of the fluctuations of the order
parameter of seismicity  that precedes a major EQ, an adequate
study of the spatial distribution of those events resulting in
this precursory minimum enables an estimate of the area in which
the major EQ will occur. We shall present applications including
the strongest EQ that occurred in California (see Fig. \ref{fig1})
during the last three decades. Analogous applications have been
completed for major EQs in Japan\cite{PNAS13,PNAS15}.

\section{The procedure for the data analysis} \label{sec2}
We first briefly describe the natural time analysis of seismicity
\cite{SPRINGER,NAT05C}. In a time series comprising $N$
earthquakes, the natural time $\chi_k = k/N$ serves as an index
for the occurrence of the $k$-th earthquake. In natural time
analysis the pair $(\chi_k, Q_k)$ is studied, where $Q_k$ is the
energy released during the $k$-th earthquake of magnitude $M_k$.
Alternatively, one may study the pair $(\chi_k,p_k)$, where $
p_k={Q_k}/{\sum_{n=1}^NQ_n}$ is the normalized energy released
during the $k$-th earthquake, and $Q_k$ -and hence $p_k$- is
estimated through the  relation \cite{KAN78} $Q_k \propto
10^{1.5M_k}$. The variance of $\chi$ weighted for $p_k$,
designated by $\kappa_1$, is given by
\cite{NAT02,SPRINGER,NAT03B,NAT03A,NAT05C}
\begin{equation}\label{kappa1}
\kappa_1=\sum_{k=1}^N p_k (\chi_k)^2- \left(\sum_{k=1}^N p_k
\chi_k \right)^2
\end{equation}
This quantity  can be also considered as an order parameter for
seismicity\cite{NAT05C,SPRINGER}.

In order to study the fluctuations of $\kappa_1$, we apply the
following procedure \cite{SPRINGER}. Taking an excerpt of a
seismic catalog comprising $W (\geq 100)$ successive events, we
start from the first EQ and  calculate the first 35 $\kappa_1$
values for 6 to 40 consecutive EQs (cf. the upper limit of this
number does not markedly influence the results \cite{NAT06B}).
Then we proceed to the second EQ, and calculate again 35 values of
$\kappa_1$ using the 7$^{th}$ to the 41$^{st}$ event. Scanning,
event by event, the whole excerpt of $W$ earthquakes, we calculate
the average value $\mu(\kappa_1)$ and the standard deviation
$\sigma(\kappa_1)$ of the $\kappa_1$ values. The quantity
\begin{equation}\label{beta}
\beta \equiv \sigma(\kappa_1)/\mu(\kappa_1)
\end{equation}
is defined\cite{NEWEPL} as the variability $\beta$ of $\kappa_1$
for this excerpt of length $W$. Usually we are interested on what
happens to the  $\beta$ value until just before the occurrence of
each EQ, $e_i$, in the seismic catalog. To achieve this goal, we
calculate first the $\kappa_1$ values using the {\em previous} 6
to 40 (or 6 to $W$)\cite{PNAS13,PNAS15} consecutive EQs. These 35 $\kappa_1$ values are associated
with the EQ $e_i$, but we clarify that the EQ $e_i$ has {\em not}
been employed for their calculation. The $\beta$ value
-corresponding to the EQ $e_i$- for a natural time window length
$W$ is computed using all the ($35 \times W$) $\kappa_1$ values
associated with the EQs $e_{i-W+1}$ to $e_i$. The resulting value
is denoted by $\beta_W$, where the subscript $W$ shows the natural
time window length, and the corresponding minimum is designated by
$\beta_{W,min}$.

We now summarize a few important points that emerged in our
earlier studies\cite{NAT05C,NAT06B}, reviewed also in chapter 6 of
ref. \onlinecite{SPRINGER}, when analyzing in natural time the
{\em long term} seismicity in wide areas. By using a natural time
window comprising 6 to 40 consecutive events sliding through an
earthquake catalog, the computed $\kappa_1$ values lead to the
construction of the probability density function (pdf)
P($\kappa_1$) of a wide area. The scaled distribution $\sigma(
\kappa_1)$ P($\kappa_1$) plotted versus ($\mu(\kappa_1) - \kappa_1
/ \sigma( \kappa_1)$) of this area collapses on the same curve
with the ones deduced from different wide areas. For example, the
curve obtained from the area $N_{25}^{46} E_{125}^{146}$ in Japan
using the JMA catalog, and the area $N_{32}^{37} W_{114}^{122}$ in
California using the southern California earthquake catalog (SCEC)
\cite{NAT05C}. It has been also found that this {\em universal}
curve of long term seismicity exhibits strikingly similar features
with the corresponding scaled distribution curves of the order
parameter fluctuations in other -nonequilibrium or equilibrium-
critical systems\cite{BRA98,BRA01,ZHE01,ZHE03,CLU04} including the
data\cite{AEG04B,KingaPRE} we analyzed\cite{NAT11A} for the   case
of self organized criticality\cite{BAK87}. This similarity was
focused on the existence of a common ``exponential tail'' (which
has in fact a double exponential form) reflecting non-Gaussian
(large, but rare) fluctuations.

\section{The data analyzed} \label{sec3}
In this study we  present examples of major EQs that occurred in
California. We used the United States Geological Survey Northern
California Seismic Network catalog available from the Northern
California Earthquake Data Center, at the http address: {\tt
www.ncedc.org/ncedc/catalog-search.html}, hereafter called NCEDC.
The earthquake magnitudes reported in this catalog are hereafter
labelled with $M$. We considered all earthquakes with $M \geq 2.5$
reported by NCEDC, within the area ${\rm N}_{31.7}^{45.7} {\rm
W}_{127.5}^{112.1}$ (the green rectangle in Fig. \ref{fig1}). We
have on average $\sim 10^2$ EQs per month since 31832 earthquakes
occurred for the 25 year period from 1 January 1979 to 1 January
2004.

Following ref. \onlinecite{EPL12},  we focus hereafter on the
natural time window length $W=$300, which is compatible with the
fact that the lead time of SES activities  is around a few months
(i.e., from 21 days to an upper limit of around 5 months, e.g.,
see ref. \onlinecite{SPRINGER}). By analyzing in natural time the
aforementioned NCEDC data, we recently identified \cite{EPL12},
for $W$=300 events, the dates of the minima $\beta_{W,min}$ as
being 1 to 5 months before four out of the five mainshocks with $M
\geq 7.0$ that occurred within the wide area  $N_{31.7}^{45.7}
W_{127.5}^{112.1}$ during the 25 year period 1 January 1979 to 1
January 2004 in California. For the two examples of the $M$7.4
Landers EQ on 28 June 1992 and the $M$7.0 Hector Mine EQ on 16
October 1999 that will be treated below, the dates of the
precursory minima $\beta_{300,min}$ were reported to be 28 January
1992 and 14 May 1999, respectively.

\section{The method proposed to estimate the epicentral area of an impending mainshock}\label{sec4}

Briefly, the method proposed in based on the following main
aspect: We argue that a minimum should be observed almost
simultaneously in the order parameter fluctuations of the
seismicity within {\em both} the wide area and the epicentral area
of the impending mainshock. This stems from our findings in ref.
\onlinecite{TECTO13} where we identified the appearance of a
$\beta_{W,min}$ (at scales $W$=100 to $W$=400) in the seismicity
within the wide area $N_{25}^{46} E_{125}^{146}$, i.e., $21^o
\times 21^o$, in Japan around the date (i.e., on 26 April 2000) of
the initiation of the SES activity recorded\cite{UYE02} at Niijima
Island -in the vicinity of which a major seismic swarm started 2
months later with a $M$6.5 EQ on 1 July 2000. Moreover, it was
found that in areas surrounding Niijima Island  the variability
$\beta$ of the seismicity exhibited a minimum almost at the {\em
same} date, i.e., around 26 April. Hence, in the light of the
above findings, the method to estimate the epicentral area of an
impending mainshock -when no information for the recording of an
SES activity is available- should consist of the following steps:

First step: We start the investigation from the natural time
analysis of the seismicity in a wide area, for example
 $N_{31.7}^{45.7} W_{127.5}^{112.1}$ (i.e., $14^o \times
15.4^o$) in California. By considering a sliding natural time
window comprising a number of events that would occur in a few
months (e.g., we may adopt, as in ref. \onlinecite{EPL12}, $W$=300
events for the aforementioned wide area in California), we search
for a minimum in the $\beta$ variability.

\begin{figure}
\begin{center}
\centerline{\includegraphics[width=.4\textwidth]{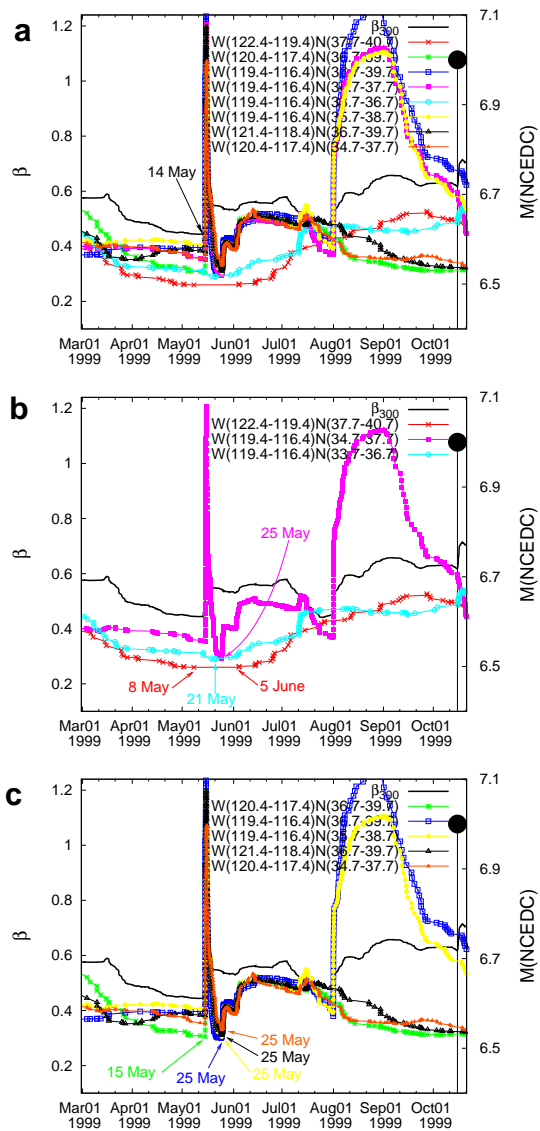}}
\caption{(color online) Results for the $M$7.0 Hector Mine EQ in
California on 16 October 1999. ({\bf a}) The variability $\beta$
of the seismicity versus the conventional time for a natural time
window length $W$ comprising a number of events that would occur
in three months: The thick black curve minimizing on 14 May 1999
corresponds to the wide area $N_{31.7}^{45.7} W_{127.5}^{112.1}$.
The other eight curves correspond to those areas -out of the 38
encircled ones in Fig. \ref{fig1}-  resulting in $\beta$ minima
which occur at dates close to that of the $\beta_{300}$ minimum of
the wide area. These dates for the lower three minima are marked
in ({\bf b}) and for the remaining five minima in ({\bf c}). ({\bf
b}): After discarding the red curve (see the text), the cyan curve
exhibits the lowest minimum on 21 May 1999. It corresponds to the
area $N_{33.7}^{36.7} W_{119.4}^{116.4}$ shown in Fig. \ref{fig1}
by the cyan rectangle along with the Hector Mine EQ epicenter
marked with the neighboring red triangle (the smaller). The EQs
with $M \geq 6.4$ (right scale) are shown with vertical bars
ending at solid circles.}\label{fig2}
\end{center}
\end{figure}

Second step: We investigate whether this minimum is a precursory
one by examining whether it obeys the criteria mentioned in ref.
\onlinecite{EPL12}. This should necessarily include the Detrended
Fluctuation Analysis (DFA)\cite{PEN94} in order to investigate
temporal correlations in the earthquake magnitude time series at
the same natural time window length scale as the one mentioned
above (cf. It has been shown \cite{JGR14} that
$\beta$ captures the correlations between the events). DFA has
become the standard method when studying long-range correlated
time series and can also be applied to real world non-stationary
signals \cite{HU01,CHE02,CHE05,MA10,XU11}.

Third step: We search for the epicentral area of the impending
mainshock by using an area window of an appreciably smaller size,
e.g., $3^o \times 3^o$ for the cases studied in California,
sliding (with steps of $1^o$ in longitude and/or $1^o$ in
latitude, see the green plus symbols in Fig. \ref{fig1}) through
the wide area. For each location of the sliding area window, we
repeat the analysis as in the wide area in order to determine the
date of $\beta_{W,min}$. Note that, if in the aforementioned wide
area of California (in which, as mentioned, occur on average 100
events/month) we used $W$=300 events,  we should consider as $W$
in each location of the sliding area window the corresponding
number of the events that would occur on average within three
months. This reflects that we cannot select an area window of
arbitrarily smaller size in view of the following limitation: In
order to identify the date of the occurrence of the minimum of
$\beta$ in an area window size with an uncertainty of around a few
days, we must have at least on the average 2 events per week,
i.e., at least 24 events for an almost three month period (12
weeks). Among these  locations of the sliding area window
investigated, we select the one(s) which led to a $\beta_{W,min}$
date which almost coincides with the date at which the
$\beta_{300,min}$ value was observed in the wide area. In
practice, the former date differs from the latter usually by no
more than a few days up to around 10 days, or so. This way we
usually find an area surrounding the future epicenter, as it
results from the examples presented in the following two Sections,
which refer to major EQs that occurred in California.

Since a major EQ occurs, as mentioned, 1 to 5 months after the
appearance of $\beta_{300,min}$ in the wide area \cite{EPL12}, the
applicability of the method proposed here becomes more clear when
a single major mainshock occurs within a five month period. Such
an example, is the case of the $M$7.0 Hector Mine EQ in California
in 1999. Furthermore, in a later Section, we shall present a case
in which two major mainshocks occurred inside the wide area within
5 months. This is the case of the $M$7.4 Landers EQ on 28 June
1992 that was preceded by the $M$6.7 ($M_s$7.1, e.g. see ref.
\onlinecite{petrolia}) Cape Mendocino EQ on 25 April 1992. As we
shall see later, it is strikingly interesting that in the latter
case the method identifies two different epicentral areas as it
should.

We clarify that in the next two Sections, we solely focus on the
third step of the method since the first two steps -referring to
the identification in the wide area of the dates of the minima
$\beta_{300,min}$ for major EQs in California and their
distinction from the non precursory minima- have been already
accomplished, as mentioned, in ref. \onlinecite{EPL12}.

To summarize: For the analysis of the examples presented below for
California, we used the wide area $N_{31.7}^{45.7}
W_{127.5}^{112.1}$  as in ref.\cite{EPL12}, and a sliding area
window size $3^o \times 3^o$ (with step $1^o$) resulting in 156
areas (see the plus symbols in Fig. \ref{fig1}) out of which 38
areas (encircled in Fig. \ref{fig1}) had at least 24 events on
average per three months. Hence, the investigation of the examples
below in California was made by analyzing the seismic data
occurring in these 38 areas.

\section{Results concerning the $M$7.0  Hector Mine EQ on 16 October 1999
at $34.60^o$N $116.34^o$W in California}\label{sec53} The thick
black curve in Fig. \ref{fig2}a depicts the variability $\beta$
versus the conventional time when a natural time window length
$W$=300 events is sliding through the wide area $N_{31.7}^{45.7}
W_{127.5}^{112.1}$. A clear minimum of $\beta$ is observed on 14
May 1999, as marked in the figure. The subsequent analysis of the
38 areas -out of the 156 depicted in Fig. \ref{fig1}- resulting
from the locations of the sliding window of size $3^o \times 3^o$
lead to 38 curves. Eight of these curves, depicted in Fig.
\ref{fig2}a, were identified to exhibit a $\beta$ minimum with
dates within 10 days, or so, from the aforementioned date on 14
May 1999 obtained from the wide area. These dates are now marked,
for reader's convenience, in two separate panels of this figure:
In Fig. \ref{fig2}b -for the three lowest minima observed in Fig.
\ref{fig2}a- and in Fig. \ref{fig2}c for the minima of the
remaining five curves. A closer inspection of Fig. \ref{fig2}b
shows that the red curve in the bottom (that is obtained from the
area $N_{37.7}^{40.7} W_{122.4}^{119.4}$) has a plateau extending
from 8 May until 5 June which in fact comes from the absence of
the occurrence of any event in the area during this almost one
month period. Thus, this curve should be discarded from further
analysis since we cannot derive a definite conclusion on the
existence or not of a minimum close to the aforementioned date 14
May 1999. Between the remaining two curves, the cyan has a minimum
which is somewhat lower and occurs on 21 May 1999. It corresponds
to the area $N_{33.7}^{36.7} W_{119.4}^{116.4}$ depicted with the
cyan rectangle in Fig. \ref{fig1} along with the epicenter of the
Hector Mine EQ (the smaller red triangle) lying at $34.60^o$N
$116.34^o$W. The epicenter is located just at the eastern edge of
the area, see the cyan rectangle in Fig. \ref{fig1}. The other
curve, i.e., the magenta in Fig.\ref{fig2}b, corresponds to the
area $N_{34.7}^{37.7} W_{119.4}^{116.4}$, see the magenta
rectangle in Fig. \ref{fig1}, which is only $0.1^o$ far from the
Hector Mine EQ epicenter. As for the five curves in
Fig.\ref{fig2}c they correspond to the areas $N_{34.7}^{37.7}
W_{120.4}^{117.4}$, $N_{36.7}^{39.7} W_{121.4}^{118.4}$,
$N_{35.7}^{38.7} W_{119.4}^{116.4}$, $N_{36.7}^{39.7}
W_{119.4}^{116.4}$ and $N_{36.7}^{39.7} W_{120.4}^{117.4}$ lying
at various distances (roughly between $1^o$ and $2.9^o$) from the
Hector Mine EQ epicenter. None of these areas, is adjacent to the
epicenter as the aforementioned one $N_{33.7}^{36.7}
W_{119.4}^{116.4}$ which corresponds to the cyan curve exhibiting
the lowest $\beta$ minimum in Figs. \ref{fig2}a,b.

Thus, in short, the cyan curve exhibiting a $\beta$ minimum on 21
May, i.e., around a week later from that (14 May) identified in
the wide area (thick black curve), which happens to be the lowest
one among the other curves showing also minima at neighboring
dates, corresponds to an area whose eastern edge lies just at the
Hector Mine EQ epicenter.

\section{Results concerning the $M$7.4 Landers EQ in California on 28 June 1992 at
$34.19^o$N $116.46^o$W}\label{sec62} 
\begin{figure*}
\begin{center}
\centerline{\includegraphics[width=0.5\textwidth]{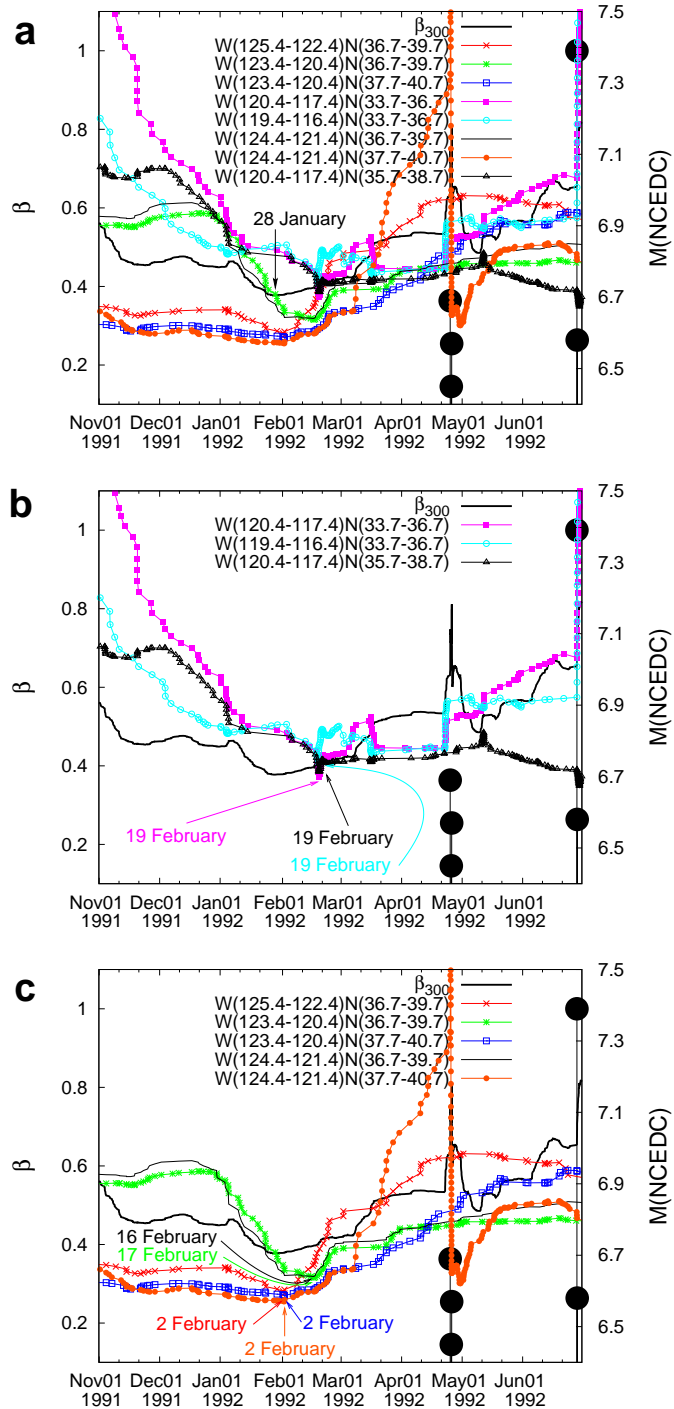}}
\caption{(color online) Results for the $M$7.4 Landers EQ in
California on 28 June 1992, and for the Cape Mendocino EQ on 25
April 1992. ({\bf a}) The variability $\beta$ of the seismicity
versus the conventional time for a natural time window length $W$
comprising a number of events that would occur in three months:
The black thick curve resulting from the wide area
$N_{31.7}^{45.7} W_{127.5}^{112.1}$ exhibits a very broad minimum
approximately from 18 January to 20 February 1992 with its lowest
value on 28 January 1992, as marked. The other eight curves
correspond to those areas -among the 38 encircled ones in Fig.
\ref{fig1}- leading to $\beta$ minima at dates within the range of
the broad minimum of the wide area. These dates are marked in
({\bf b}) for the three curves that correspond to areas related
with the Landers EQ and in ({\bf c}) for the remaining five curves
corresponding to areas associated with the Cape Mendocino EQ (see
the text). The cyan curve in ({\bf b}) corresponds to the area
$N_{33.7}^{36.7} W_{119.4}^{116.4}$ depicted in Fig. \ref{fig1}
with the cyan rectangle that includes the Landers EQ epicenter
(the largest red triangle). The curve with the orange circles in
({\bf c}) corresponds to the area $N_{37.7}^{40.7}
W_{124.4}^{121.4}$, shown in Fig. \ref{fig1} with the orange
rectangle surrounding the Cape Mendocino EQ epicenter. The EQs
with $M \geq 6.4$ (right scale) are shown with vertical bars
ending at solid circles.}\label{fig3}
\end{center}
\end{figure*}

This EQ, which is the largest
moment magnitude ($M_w$) EQ that occurred in California since
1992, has been preceded by the $M$6.7 ($M_s$7.1) Cape Mendocino EQ
on 25 April 1992 with an epicenter at $40.36^o$N $124.23^o$W
(followed by two strong $M$6.45 and $M$6.6 aftershocks on 26 April
1992). Hence, Landers EQ and Cape Mendocino EQ occurred within
almost 3 months being less than the maximum lead time between
$\beta_{300,min}$ and the subsequent EQ which is around 5 months
\cite{EPL12}. The complexity of this case might be the origin of
the following fact: The variability of the seismicity versus the
conventional time in the wide area $N_{31.7}^{45.7}
W_{127.5}^{112.1}$, depicted  in Fig. \ref{fig3} with a thick
black curve, although exhibited its lowest value on 28 January
1992 (marked in Fig. \ref{fig3}a, its minimum is in fact very
broad (like a plateau) lasting for about 4 weeks, i.e., from 18
January to around 20 February, 1992. This must be carefully
considered when the area window of $3^o \times 3^o$ is sliding
(with step $1^o$) through the aforementioned wide area. We found
that, among the 38 areas studied,  eight of them led to $\beta$
minima with dates from 2 February  to 19 February 1992, thus lying
within the broad range mentioned above for the $\beta$ minimum of
the wide area. The eight curves corresponding to these areas are
plotted in Fig. \ref{fig3}a and can be classified into two types,
which are depicted separately in Figs. \ref{fig3}b and \ref{fig3}c
where we also mark the date of the minimum in each curve. First,
in three of these curves shown with magenta squares, black
triangles and cyan circles, which correspond to the areas
$N_{33.7}^{36.7} W_{120.4}^{117.4}$, $N_{35.7}^{38.7}
W_{120.4}^{117.4}$ and $N_{33.7}^{36.7} W_{119.4}^{116.4}$, a very
sharp minimum appears on 19 February 1999, see Fig. \ref{fig3}b.
An inspection of these three areas shows that the epicenter of
Landers EQ in Fig. \ref{fig1} (large red triangle) lies inside the
latter area (depicted with a cyan rectangle in Fig. \ref{fig1}),
$1^o$ to the east of the first area and around $1.8^o$ to the
southeast of the second area. Second, the remaining five curves,
see Fig. \ref{fig3}c, shown in orange circles, blue squares, red
crosses, thin black line and green asterisks (starting from the
curve with the lowest $\beta$ minimum), exhibit non-sharp minima
on the following dates 2, 2, 2, 16 and 17 February, 1992, and
correspond to the following five areas: $N_{37.7}^{40.7}
W_{124.4}^{121.4}$, $N_{37.7}^{40.7} W_{123.4}^{120.4}$,
$N_{36.7}^{39.7} W_{125.4}^{122.4}$, $N_{36.7}^{39.7}
W_{124.4}^{121.4}$ and $N_{36.7}^{39.7} W_{123.4}^{120.4}$,
respectively. The first area is depicted with the orange rectangle
in Fig.\ref{fig1}, where we see that the epicenter of the Cape
Mendocino EQ marked by a red triangle -inside the orange
rectangle- along with the neighboring two smaller red triangles
showing its two aftershocks\cite{petrolia}, lies inside it (cf.
the brown rectangle depicted in Fig, \ref{fig1} also results as a
candidate epicentral area for the Cape Mendocino EQ upon
considering a reasonable experimental error in the determination
of the magnitudes.
Concerning the other areas, we find that the Cape Mendocino
epicenter lies $0.8^o$ to the West of the second area, $0.8^o$ to
the North from the third and the fourth, and around $1.1^o$ to the
NW of the fifth area.

Let us summarize: We found that eight curves  minimize at dates
within the broad minimum (18 January - 20 February) exhibiting by
the thick black curve corresponding to the wide area. Among these
curves, three minimize on 19 February 1992 and five at earlier
dates. The former three curves correspond to areas related with
Landers EQ since one of them is surrounding its epicenter, and the
other two are neighboring areas, i.e., $1^o$ and $1.8^o$ far from
the epicenter. The remaining five curves correspond to areas
related with the Cape Mendocino EQ since one of them (exhibiting
the lowest minimum) is surrounding its epicenter, and the other
four lie in its vicinity.

\section{Discussion and Conclusions}\label{sec7}
The prediction of the epicenter of an impending mainshock can be
made on the basis of the SES physical properties, as mentioned in
the Introduction. This of course has two prerequisites: First,
that geoelectrical data are available and second, that we know
from earlier experience the selectivity map \cite{VAR96B} of the
measuring station at which the SES under investigation appeared.
The method proposed here cannot be falsely considered as replacing
that followed by means of SES since it does not improve its
accuracy, which is of the order of 100km, but it has the privilege
that can be applied even in the lack of geoelectrical data.

In conclusion, here we proposed a method to estimate the
epicentral area of an impending mainshock based on a precursory
minimum observed in the fluctuations of the order parameter of
seismicity. This has been successfully applied to mainshocks in
California including the  Landers EQ on 28 June 1992, which is the
largest  earthquake that occurred in California during the last
three decades. These results are found to be robust when repeating 100 trials upon
considering a reasonable experimental error in the magnitude
measurements of the earthquakes in the seismic catalog used.

\section{Methods}
All the seismic data taken from NCEDC have been analyzed in
natural time. The seismic moment $M_0$, which is proportional to
the energy release during an earthquake and hence to the quantity
$Q_k$ used in natural time analysis, is calculated \cite{SPRINGER}
from the relation $\log_{10}(M_0)=1.5M+$const.

%
%





\end{document}